\documentclass[aps,pra,twocolumn,groupedaddress,showpacs]{revtex4}
\usepackage{bm}
\usepackage{epsf}
\usepackage{amssymb}
\usepackage{amsmath}
\usepackage{graphicx}
\usepackage{rotating}
\usepackage{epsfig}
\usepackage{psfrag}
\usepackage{amsmath}
\usepackage{hyperref}

\hypersetup{
    bookmarks=true,         
    unicode=false,          
    pdftoolbar=true,        
    pdfmenubar=true,        
    pdffitwindow=true,      
    pdftitle={My title},    
    pdfauthor={Author},     
    pdfsubject={Subject},   
    pdfcreator={Creator},   
    pdfproducer={Producer}, 
    pdfkeywords={keywords}, 
    pdfnewwindow=true,      
    colorlinks=false,       
    linkcolor=red,          
    citecolor=green,        
    filecolor=magenta,      
    urlcolor=cyan           
}

\DeclareMathAlphabet{\bi}{OML}{cmm}{b}{it}

\begin{document}
\def\G{{\cal G}}
\def\F{{\cal F}}
\def\ea{\textit{et al.}}
\def\bM{{\bm M}}
\def\bN{{\bm N}}
\def\bV{{\bm V}}
\def\bj{\bm{j}}
\def\bSig{{\bm \Sigma}}
\def\bLam{{\bm \Lambda}}
\def\bfeta{{\bf \eta}}
\def\bc{{\bf c}}
\def\ba{{\bf a}}
\def\d{{\bf d}}
\def \xy{$x$--$y$ }
\def\bP{{\bf P}}
\def\bK{{\bf K}}
\def\bk{{\bf k}}
\def\bkn{{\bf k}_{0}}
\def\bx{{\bf x}}
\def\bz{{\bf z}}
\def\bR{{\bf R}}
\def\br{{\bf r}}
\def\bu{{\bm u}}
\def\bq{{\bf q}}
\def\bp{{\bf p}}
\def\bG{{\bf G}}
\def\bQ{{\bf Q}}
\def\bs{{\bf s}}
\def\bA{{\mathbf A}}
\def\bv{{\bf v}}
\def\b0{{\bf 0}}
\def\la{\langle}
\def\ra{\rangle}
\def\Im{\mathrm {Im}\;}
\def\Re{\mathrm {Re}\;}
\def\beq{\begin{equation}}
\def\eeq{\end{equation}}
\def\bea{\begin{eqnarray}}
\def\eea{\end{eqnarray}}
\def\bdm{\begin{displaymath}}
\def\edm{\end{displaymath}}
\def\bnab{{\bm \nabla}}
\def\Tr{{\mathrm{Tr}}}
\def\bJ{{\bf J}}
\def\bU{{\bf U}}
\def\bPsi{{\bm \delta\Delta}}
\def\mA {\mathrm{A}}
\def \R{R_{\mathrm{s}}}
\def \rhos{n_{\mathrm{s}}}
\def \rhon{\tilde{n}}
\def \Rd{R_{\mathrm{d}}}
\def \xy{three dimensional $XY\;$}
\def\sfrac{\textstyle\frac}

\title{Critical Behavior in Trapped Strongly Interacting Fermi Gases}
\author{E.~Taylor}
\affiliation{Dipartimento~di~Fisica and CNR-INFM BEC center, Universit\`a di Trento, I-38050 Povo, Trento, Italy}
\altaffiliation{Present address: Department of Physics, The Ohio State University, Columbus, Ohio, 43210}

\date{Aug. 20, 2009}

\begin{abstract}
We investigate the width of the Ginzburg critical region and experimental signatures of critical behavior in strongly interacting trapped Fermi gases close to unitarity, where the $s$-wave scattering length diverges.   Despite the fact that the width of the critical region is of the order unity, evidence of critical behavior in the bulk thermodynamics of trapped gases is strongly suppressed by their inhomogeneity.  The specific heat of a harmonically confined gas, for instance,  is \textit{linear} in the reduced temperature $t = (T-T_{\mathrm{c}})/T_{\mathrm{c}}$ above $T_{\mathrm{c}}$.  We also discuss the prospects of observing critical behavior in the local compressibility from measurements of the density profile.
\end{abstract}
\maketitle

\section{Introduction}

Atomic Fermi gases close to unitarity~\cite{Trentoreview} are among the most strongly interacting systems known.  Understanding their thermodynamic properties is an important theoretical~\cite{Chen05,Haussmann07,Hu07,Bulgac07,Haussmann08} and experimental{~\cite{Kinast05,Luo07,Luo08} challenge that is only beginning to be addressed.   In typical strongly interacting superfluids such as $^4$He,  thermodynamic measurements at low temperatures outside the critical region reveal the nature of the elementary quasiparticles (e.g., phonons and rotons)~\cite{LPSM}.  Conversely, in the critical region close to the superfluid transition temperature $T_{\mathrm{c}}$, phase fluctuations of the order parameter play a crucial role and simple theories based on this quasiparticle picture breakdown.  In this region, the temperature dependencies of thermodynamic quantities are instead governed by universal scaling laws~\cite{Zinn}.   
The width of the critical region reflects the strength of interactions in the system: for weak-coupling BCS superconductors, it is very small, with magnitude $(T_{\mathrm{c}}/\epsilon_{\mathrm{F}})^4 \sim 10^{-14}$~\cite{Kadanoff}.  In contrast, for strongly interacting clean high-$T_{\mathrm{c}}$ cuprate superconductors, it can reach  $10^{-2}$~\cite{Cobb,Deutscher}.  In trapped, weakly interacting atomic Bose gases, it is typically between $10^{-4}$ and $10^{-2}$~\cite{Giorgini96,Damle96,Donner07}.   One of the widest critical regions belongs to superfluid $^4$He, where it is of the order unity.  The existence of such a large critical region gives rise to a clear signature of critical behavior in thermodynamic quantities, most dramatically in the specific heat which exhibits the famous lambda curve~\cite{Heexpt}.

An important question in the study of trapped strongly interacting Fermi gases is to what extent the measured temperature dependencies of thermodynamic quantities~\cite{Kinast05,Luo07,Luo08} reflect a gas of weakly interacting quasiparticles as opposed to universal scaling behavior arising from phase fluctuations in the critical region.   In contrast to the situation in trapped weakly interacting Bose gases,  the strong interactions in Fermi gases close to unitarity can give rise to a large critical region, suggesting that scaling behavior (of the \xy universality class) might be evident in bulk thermodynamic quantities such as the specific heat. Interpreting thermodynamic measurements in trapped gases is complicated by the inhomogeneity of the gas, however, since the superfluid transition temperature $T_{\mathrm{c}}(r)$ depends on the distance $r$ from the center of the trapping potential (assumed to be isotropic).  This means that sharp signatures of critical behavior in bulk thermodynamic quantities that exist in homogeneous samples become smoothed out by the averaging effects of the nonuniform density profile in the trap.  

Our aim in this paper is to understand whether or not current experiments with trapped Fermi gases are observing signatures of critical behavior and also to act as a guide for future experiments looking for such signatures.  Building on the work of  S\'a de Melo, Randeria, and Engelbrecht~\cite{SadeMelo93}, in Sec.~\ref{GCRsec}, we study the width of the critical region of a two-component Fermi gas through the BCS-BEC crossover~\cite{BCSBEC} using the microscopic theory of Nozi\`eres and Schmitt-Rink (NSR)~\cite{NSR}.   That the critical region in a Fermi gas close to unitarity should have a width of order unity has been pointed out by Randeria~\cite{RanderiaTalk}, based on the fact that the $T=0$ Ginzburg--Landau coherence length is very small there, comparable to the mean inter-particle spacing $k^{-1}_{\mathrm{F}}$~\cite{Pistolesi94,Engelbrecht97}.  We confirm that the critical region has a width of order unity in the strongly interacting region close to unitarity, reaching a maximum just on the BEC side of resonance.  

In Sec.~\ref{SHsec}, we derive the universal scaling relation for the trap averaged bulk specific heat $C = \int d\br c(r)$ in a Fermi gas confined by a power-law trapping potential.  We also discuss the scaling properties of the local isothermal compressibility $\kappa_T(r)$, since this quantity can be measured by looking at the density profile $n(r)$ of the trapped gas.  

The effect of a harmonic trapping potential on bulk thermodynamic quantities in a Bose gas has been investigated by Damle \textit{et al.} in Ref.~\cite{Damle96}.  Using a local density approximation, they calculate the scaling behavior of various quantities by substituting the local transition temperature $T_{\mathrm{c}}(r)$ into well-known scaling relations developed for homogeneous \xy systems.   This approach has also been used to determine the specific heat critical exponent $\alpha$ in high-precision measurements~\cite{Heexpt2,Ahlers1,Ahlers2} performed on $^4$He in large sample sizes, where gravity effects produce a position-dependent transition temperature as well.  In this paper, we carry out a detailed analysis of the bulk specific heat of a strongly interacting trapped Fermi gas.  Despite the large critical region, we find that the critical scaling behavior of the bulk specific heat in a trapped Fermi gas is strongly suppressed relative to that in a uniform superfluid.  

Finally, in Sec.~\ref{Implicationssec}, we discuss the relevance of our results to current experiments on strongly-interacting Fermi gases.

\section{Ginzburg Critical Region}
\label{GCRsec}
The Ginzburg critical region defines the range of temperatures about $T_{\mathrm{c}}$ where the root mean square thermal fluctuations of the order parameter exceed some characteristic mean-field value~\cite{LLSM}. In the superfluid state, this characteristic value is the mean-field order parameter while an analogous value can be defined in the normal phase.  The width of the critical region is estimated from the Ginzburg--Landau functional~\cite{LLSM}
\beq \Delta\Omega = \int d\br\left[A|\delta\Delta|^2 + \frac{B}{2}|\delta\Delta|^4 + D|\bnab\delta\Delta|^2\right]\label{GL}\eeq
for fluctuations $\delta\Delta$ of the order parameter $\Delta_0$.  (This expression is valid for the normal phase, where $\Delta_0=0$.  A similar expression describes the fluctuations in the superfluid phase~\cite{LLSM}.) Here, $A(T) = a t$, where $t\equiv (T-T_{\mathrm{c}})/T_{\mathrm{c}}$.  In the vicinity of $T_{\mathrm{c}}$, the root mean square thermal fluctuations of the order parameter are much smaller than the characteristic value $\sqrt{2at/B}$ as long as $t\gg Gi$, where~\cite{noteGL} 
\beq Gi \equiv \frac{B^2k^2_{\mathrm{B}}T^2_{\mathrm{c}}}{32\pi^2a D^3}\label{Gi}\eeq 
is the Ginzburg--Levanyuk number.  For values of $T$ inside the critical region ($|t|<Gi$), fluctuations are large, meaning that the mean-field Landau theory of phase transitions breaks down and critical behavior sets in.

In the remainder of this Section, we calculate the width ($Gi$) of the critical region through the BCS-BEC crossover~\cite{BCSBEC,NSR} for Fermi superfluids using NSR theory.  For a two-component gas of fermions, tuning the $s$-wave scattering length $a_s$, one can access the entire crossover, from a weak-coupling BCS superfluid of Cooper pairs ($a_s$ small and negative) to a Bose-condensate of tightly-bound dimer molecules in the BEC limit ($a_s$ small and positive).  In between, the scattering length diverges to infinity at unitarity.  The region of strong interactions close to unitarity has been realized in experiments on trapped Fermi gases with Feshbach scattering resonances~\cite{Trentoreview}.

\begin{figure}
\begin{center}
\epsfig{file=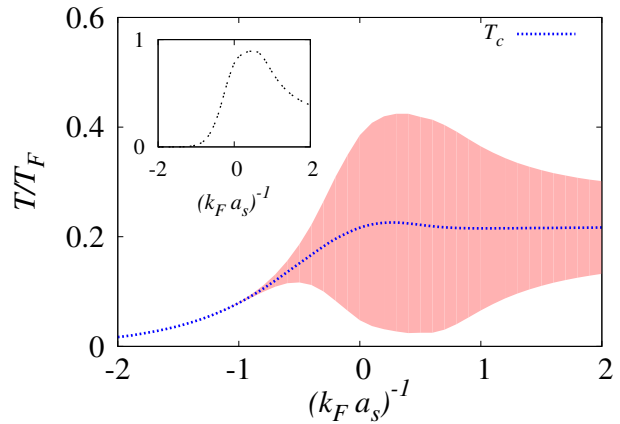, angle=0,width=0.47\textwidth}
\caption{(Color online) Ginzburg critical region through the BCS-BEC crossover.  The dashed line denotes the critical temperature $T_{\mathrm{c}}$ calculated within the NSR formalism as a function of the dimensionless interaction parameter $(k_{\mathrm{F}}a_s)^{-1}$ from the BCS region ($(k_{\mathrm{F}}a_s)^{-1}< 0$) into the BEC region ($(k_{\mathrm{F}}a_s)^{-1}>0$) (from the calculation in Ref.~\cite{Fukushima07}).  The shaded region is the critical region defined by $|T-T_{\mathrm{c}}|/T_{\mathrm{c}} < Gi$.  Inset: the Ginzburg--Levanyuk number $Gi$.}
\label{tkappa}
\end{center}
\end{figure}

NSR theory is the simplest theory of the BCS-BEC crossover that gets the correct superfluid transition temperature in the BEC limit of strongly-coupled dimer molecules.  In contrast to mean-field BCS theory, which predicts a transition temperature equal to the disassociation temperature of dimer molecules, NSR gives the temperature at which a gas of dimers Bose-condenses~\cite{NSR,SadeMelo93}, $~\sim 0.21T_{\mathrm{F}}$ (see Fig.~\ref{tkappa}), where $T_{\mathrm{F}}\equiv \epsilon_{\mathrm{F}}/k_{\mathrm{B}}$ is the Fermi temperature and $\epsilon_{\mathrm{F}}$ is the Fermi energy).  We emphasize, however, that NSR is still ``mean-field" insofar as it is a Ginzburg--Landau expansion of the thermodynamic potential (action) in powers of order parameter fluctuations close to $T_{\mathrm{c}}$:~\cite{SadeMelo93}
\beq \frac{S}{\beta} \!\simeq \!\sum_q|\delta\Delta_q|^2\Gamma^{-1}_q\!+\!\frac{U_0}{2}\!\sum_{qpp'}\!\delta\Delta^*_{p+q}\delta\Delta^*_{p'-q}\delta\Delta_{p}\delta\Delta_{p'}\!+\!\cdots.\label{S} \eeq
Here, $q \equiv (\bq,iq_m)$, $p$, and $p'$ are momentum/Matsubara frequency four-vectors, 
\beq \Gamma^{-1}_q=
\sum_{\bk}\left[\frac{1-f_{\bk} - f_{\bk-\bq}}{iq_m - \xi_{\bk} - \xi_{\bk-\bq}} + \frac{1}{2\epsilon_{\bk}}\right]-\frac{m}{4\pi a_s} \eeq
is the inverse two-particle vertex function for fermions with $s$-wave scattering length $a_s$, $f_{\bk} = [\exp(\beta\xi_{\bk})+1]^{-1}$ is the Fermi thermal distribution, $\xi_{\bk} = \epsilon_{\bk}-\mu$, and $\epsilon_{\bk}=\bk^2/2m$.  We set $\hbar=1$ throughout this paper, unless stated otherwise. $U_0$ is the long-wavelength, static limit of the effective interaction between Cooper pairs:
\beq U_0 = \sum_{\bk}\left[\frac{X}{4\xi^3_{\bk}} -\frac{Y}{8\xi^2_{\bk}}\right]. \eeq
Here, $X \equiv \tanh(\beta\xi_{\bk}/2)$ and $Y\equiv \beta \mathrm{sech}^2(\beta \xi_{\bk}/2)$.  
$T_{\mathrm{c}}$  is determined from $\Gamma^{-1}_{0}|_{T=T_{\mathrm{c}}}=0$, where the chemical potential $\mu$ is calculated from the number equation~\cite{SadeMelo93} $n = k^3_{\mathrm{F}}/3\pi^2 = \sum_{\bk}(1-X) + \beta^{-1}\sum_q\Gamma^{-1}_q\partial\Gamma_q/\partial\mu$, where $k_{\mathrm{F}}$ is the Fermi wavevector.

The coefficients in the \textit{classical} Ginzburg--Landau functional, (\ref{GL}), can be obtained from the static limit ($q_m=0$) of the quantum action, (\ref{S}), using $\Delta\Omega = S/\beta$.  They are given by $ D = \frac{1}{2}(\partial^2\Gamma^{-1}_q/\partial \bq^2)|_{q=0,T=T_{\mathrm{c}}}$, $ a = T_{\mathrm{c}}(\partial \Gamma^{-1}_{0}/\partial T)|_{T=T_{\mathrm{c}}}$, and $B = U_0|_{T=T_{\mathrm{c}}}$.  Explicitly,
\beq D \!=\! \frac{1}{2m}\sum_{\bk}\left[\frac{X}{4\xi^2_{\bk}}\! -\! \frac{Y}{8\xi_{\bk}}\!+\!\frac{k^2_z}{8m\xi_{\bk}}\!\left(\!\!XY\beta_{\mathrm{c}} \! +\! \frac{Y}{\xi_{\bk}} \!-\! \frac{2X}{\xi^2_{\bk}}\!\right)\!\right]
\eeq
and
\beq a\! =\frac{1}{4}\sum_{\bk}Y-T_{\mathrm{c}}\left(\frac{\partial\mu}{\partial T}\right)\sum_{\bk}\left[\frac{X}{2\xi^2_{\bk}} -\frac{Y}{8\xi_{\bk}}\right], 
\label{A}\eeq
where $\beta_{\mathrm{c}} \equiv 1/k_{\mathrm{B}}T_{\mathrm{c}}$ and all terms are evaluated at $T=T_{\mathrm{c}}$.  

There is no analog of the second term in (\ref{A}) (involving the temperature derivative of the chemical potential) in the classic analysis of Gork'ov~\cite{Gorkov}, which considered the BCS limit of the BCS-BEC crossover.  In the BCS limit, this term is negligible.  However, it becomes increasingly important through the crossover, finally becoming the dominant term in the BEC limit.  In this limit, where $a_s>0$ is small and positive, the NSR theory of a two-component Fermi gas reduces to the Bogoliubov--Popov theory~\cite{ShiGriffin} of a dilute Bose gas of dimer molecules with mass $M=2m$~\cite{Andrenacci03,Pieri05}. 
In this theory, the Bose chemical potential $\mu_B$ vanishes at $T_{\mathrm{c}}$ and, constructing a Ginzburg--Landau theory based on it, we would have $A =- \mu_B$ and $a =- T_{\mathrm{c}}(\partial\mu_B/\partial T)_{T=T_{\mathrm{c}}}$.  
In the BEC limit of NSR theory, $\Gamma^{-1}_q \rightarrow {\cal{Z}}(-iq_m + \bq^2/4m - \mu_B)$, where ${\cal{Z}}\equiv m^2a_s/8\pi$ and $\mu_B = 2\mu + |E_b|$ is the chemical potential for dimer molecules with binding energy $E_b = -1/ma^2_s$~\cite{Andrenacci03,Pieri05}.  The Thouless criterion $\Gamma^{-1}_0|_{T=T_{\mathrm{c}}}=0$ for Fermi superfluids thus  also requires $\mu_B$ to vanish at $T_{\mathrm{c}}$ and consequently, $a  =-{\cal{Z}}T_{\mathrm{c}}(\partial \mu_B/\partial T)_{T=T_{\mathrm{c}}}$, consistent with the Bogoliubov--Popov theory of dimer molecules.  Thus,  we see how the second term in (\ref{A}) is crucial for obtaining the correct behavior of $a$ in the BEC limit.

In the BEC limit, the chemical potential $\mu \simeq E_b/2 \ll 0$ is approximately one-half the binding energy of a dimer molecule and one finds $D = m a_s/32\pi$, $B = m^3a^3_s/16\pi$, and $a = -[T_{\mathrm{c}}(\partial\mu/\partial T)|_{T=T_{\mathrm{c}}}]m^2a_s/4\pi$.  Expanding the BCS gap equation in powers of $\Delta_0/|\mu|$, one obtains (see also Eq.~(38) in Ref.~\cite{Diener08}) $\mu_B \simeq U_{dd} n_c(T)$ in the superfluid phase, where $U_{dd} = 4\pi a_s/m$ is the mean-field dimer-dimer interaction and~\cite{Fukushima07} $n_c(T) = \Delta^2_0(T) m^2 a_s/8\pi$ is the condensate density of dimers.  Using the approximation $\Delta_0(T) \approx \Delta_0(0)\sqrt{-t}$ with $\Delta_0(0) = 4\epsilon_{\mathrm{F}}/\sqrt{3\pi k_{\mathrm{F}}a_s}$, one finds $a = m k^3_{\mathrm{F}} a^2_s/12\pi^2$ and $Gi =12\pi (k_{\mathrm{F}}a_s)(k_{\mathrm{B}}T_{\mathrm{c}}/\epsilon_{\mathrm{F}})^2~\sim(k_{\mathrm{F}}a_s)$.  Note that this coincides with a known condition $|t|\lesssim n^{1/3}_da_{dd}$ for the Bogoliubov--Popov theory of a dilute Bose gas (here, a gas of dimers with density $n_d=n/2$ and scattering length $a_{dd}=2a_s$) to break down~\cite{ShiGriffin}.  

In the BCS limit where $a_s<0$ is small and $\mu \simeq \epsilon_{\mathrm{F}}$, it is straightforward to show (see the related calculations in Ref.~\cite{SadeMelo93}) that $a = N(\epsilon_{\mathrm{F}})$, $D = 7\zeta(3)N(\epsilon_{\mathrm{F}})\epsilon_{\mathrm{F}}/24\pi^2m(k_{\mathrm{B}}T_{\mathrm{c}})^2$, and $B=7\zeta(3)N(\epsilon_{\mathrm{F}})/8\pi^2(k_{\mathrm{B}}T_{\mathrm{c}})^2$, where $N(\epsilon_{\mathrm{F}})$ is the density of states at the Fermi surface.  Thus, the width of the critical region is proportional to $(T_{\mathrm{c}}/\epsilon_{\mathrm{F}})^4$, as expected.

In the region close to unitarity, the GL coefficients are determined numerically using the NSR values for the chemical potential at $T_{\mathrm{c}}$~\cite{Fukushima07}.  To avoid errors associated with taking a numerical derivative of the chemical potential with respect to temperature (in fact, the temperature dependence of the NSR chemical potential exhibits unphysical features in the critical region~\cite{Fukushima07}), we take the temperature derivative of the gap equation.  This gives us an equation for $(\partial\mu/\partial T)|_{T=T_{\mathrm{c}}}$ in terms of $\mu_{\mathrm{c}}$, $T_{\mathrm{c}}$, and $(\partial\Delta_0/\partial T)_{T=T_{\mathrm{c}}}$ in the superfluid phase. (Strictly speaking, the Ginzburg--Levanyuk number will assume different values in the normal and superfluid phase.  This difference is not crucial, however, and we take $Gi$ to be symmetric.)  For this last quantity, we use  $\Delta_0(T) \approx \Delta_0(0)\sqrt{-t}$, where $\Delta_0(0)$ is taken from Ref.~\cite{Fukushima07}.  We plot the critical region through the BCS-BEC crossover in Fig.~\ref{tkappa}.  The Ginzburg--Levanyuk number is very small in the BCS region ($(k_{\mathrm{F}}a_s)^{-1}<0$), approaches unity close to unitarity, and slowly shrinks again in the BEC region ($(k_{\mathrm{F}}a_s)^{-1}>0$).  It is interesting to note that the critical region is actually widest on the BEC side of resonance, close to the point where the chemical potential vanishes.  Related nonmonotonic dependence of the Ginzburg--Landau coherence length was found in Ref.~\cite{SadeMelo93}.

For a trapped gas, the fact that the Ginzburg critical region close to unitarity is of order unity means that when the temperature at the center of the trap reaches the local superfluid transition $T_{\mathrm{c}}(r=0)$, a significant fraction of the gas will be in the critical region.  In the critical region, thermodynamic quantities should exhibit universal power-law scaling, with scaling laws suitably modified by the presence of the trapping potential~\cite{Damle96}.  With this in mind, we now turn our attention to properties of the specific heat and compressibility in trapped Fermi gases in the vicinity of the superfluid transition.

\section{Specific heat and compressibility in trapped gases}
\label{SHsec}

For a homogeneous \xy superfluid, thermodynamic quantities such as the specific heat $C_V$ at constant volume and the compressibility $\kappa_T$ at constant temperature exhibit simple scaling behavior near the critical point.  For the specific heat, one has~\cite{Zinn}
\bea C_V =  N k_{\mathrm{B}}\left[f|t|^{-\alpha} + g + h_c(n,T)\right].\label{cU}\eea
Here, $\alpha\sim -0.01$~\cite{Heexpt2,Guillou}, and $f$ and $g$ are dimensionless coefficients, with different values above ($f_n,g_n$) and below ($f_s,g_s$) the superfluid transition, and $h(n,T)$ describes the analytic background contributions to the specific heat.  In general, $f$ and $g$ depend on density but not temperature ((\ref{cU}) displays the asymptotic temperature dependence in the limit $T\to T_{\mathrm{c}}$).   Similarly, for the compressibility $\kappa_T\equiv n^{-2}(\partial n/\partial \mu)_T$, we have
\bea \kappa_T = \frac{1}{n\epsilon_{\mathrm{F}}(n)}\left[k|t|^{-\gamma} + h_{\kappa}(n,T)\right],\label{kU}\eea
where $\gamma \simeq 1.3$~\cite{Guillou2} and $k=k(n)$ is a dimensionless function of density that assumes different values above and below the superfluid transition.  $h_{\kappa}(n,T)$ is a dimensionless function of density and temperature describing background contributions to the compressibility.  

Equations (\ref{cU}) and (\ref{kU}) can be extended to inhomogeneous systems by making use of the local density approximation (LDA):
\bea \mu[n(r)] = \mu_0 - V_{\mathrm{ext}}(r).\label{LDA}\eea
Here, $\mu[n(r)]$ is the local chemical potential, $\mu_0$ is the equilibrium position-independent chemical potential, and $V_{\mathrm{ext}}$ is the external trapping potential.  We can use (\ref{LDA}) to obtain an expression for the local superfluid transition temperature $T_{\mathrm{c}}(r)$ since $T_{\mathrm{c}}\propto \mu_{\mathrm{c}}$, the chemical potential at $T_{\mathrm{c}}$.  Thus, for a power-law potential $V_{\mathrm{ext}}(r) = V_pr^p$, $T_{\mathrm{c}}(r) = T_{\mathrm{c}}(0)[1-x^p]$, where $x\equiv r/R_{\mathrm{TF}}$ and $R_{\mathrm{TF}}\equiv (\mu_0/V_p)^{\frac{1}{p}}$ is the Thomas--Fermi radius of the gas.  The local reduced temperature $t(r)\equiv [T-T_{\mathrm{c}}(r)]/T_{\mathrm{c}}(r)$ is
\bea t(r) = \frac{t(0)+x^p}{1-x^p},\label{Tcr}\eea
where $t(0) = [T- T_{\mathrm{c}}(0)]/T_{\mathrm{c}}(0)$ is the reduced temperature at the trap center.  When $t(0)<0$, the system becomes superfluid, with the order parameter appearing at the trap center where $t(r)$ is smallest. %

In the remainder of this Section, we apply the local density approximation (\ref{Tcr}) to (\ref{cU}) and (\ref{kU}) to determine the critical properties of the specific heat and compressibility in trapped gases.  Such an LDA approach is standard in the cold atom literature (see, for instance, Refs.~\cite{Damle96,Ho09,Zhou09}), although we emphasize that it breaks down for very small $t(0)$ when the coherence length becomes too large.  In Appendix~\ref{LDAsec}, we give a brief discussion of the validity of LDA in the critical region.  

\subsection{Specific heat in a trapped gas}
Using (\ref{Tcr}) in (\ref{cU}) and recalling that the coefficients $f$ and $g$ assume different values above and below $T_{\mathrm{c}}$, for small values of $t(0)$, the specific heat per unit volume in a trapped gas is
\bea \frac{c}{n(r) k_{\mathrm{B}}} \!=\! \left\{\!\! \begin{array}{lcr}f_s|t(r)|^{-\alpha}\! +\! g_s\!+\!h_c(n,T)\! &\mbox{for}\!
& r\!<\!R_s
\\
f_n|t(r)|^{-\alpha}\!+ \!g_n\!+\!h_c(n,T) \!&\mbox{for}
& \!R_s\!<\!r\!<\! R_{\mathrm{c}}.
\end{array}\right.\nonumber\\ 
\label{cT}\eea
Here, $R_s = R_{\mathrm{TF}}|t(0)|^{\frac{1}{p}}$ is the radius of the superfluid region.  When $T<T_{\mathrm{c}}(0)$, the system is superfluid for $r<R_s$ and normal outside.  $R_{\mathrm{c}}$ is the Ginzburg critical radius, defined by $|t(R_{\mathrm{c}}-R_s)| = Gi$.  It describes the region is space where the local reduced temperature is inside the Ginzburg critical region and we expect the scaling behavior given by (\ref{cU}) to apply.  In the normal phase, $R_s=0$ and the specific heat is given by the second line in (\ref{cT}).

Experiments~\cite{Kinast05,Luo08} measure the \textit{bulk} specific heat $C\equiv 4\pi\int^{R_{\mathrm{TF}}}_0 dr r^2 c(r)$ of a trapped gas.  Using the above results and the Thomas--Fermi expression~\cite{Trentoreview} $n(r)=n(0)[1-x^p]^{3/2}$ for the density profile, for $T<T_{\mathrm{c}}(0)$ ($t(0)<0$), we obtain the following expression for $C$:
\bea C\!\! &=&\!\!\! \int^{|t(0)|^{\frac{1}{p}}}_0\! dx x^2 \tilde{f}_s(x,T)\left[1-x^p\right]^{\frac{3}{2}+\alpha}
\left[|t(0)|-x^p\right]^{-\alpha}\nonumber\\&&\!\!\!\!\!\!\!\!\!\!+
\int^{R'_{\mathrm{c}}}_{|t(0)|^{\frac{1}{p}}}\! dx x^2 \tilde{f}_n(x,T)\left[1-x^p\right]^{\frac{3}{2}+\alpha}
\left[x^p-|t(0)|\right]^{-\alpha}\nonumber\\&&+H^{-}(T).
\label{Capprox}\eea
For $T>T_{\mathrm{c}}(0)$ ($t(0)>0$), the bulk specific heat becomes
\bea C \!\! &=&\!\! \!
\int^{R'_{\mathrm{c}}}_{0} \!x^2 \tilde{f}_n(x,T)\left[1-x^p\right]^{\frac{3}{2}+\alpha}
\left[x^p+t(0)\right]^{-\alpha}\!\!+\!H^{+}(T).\nonumber\\
\label{Capprox2}\eea
In these equations, $R'_{\mathrm{c}}\equiv R_{\mathrm{c}}/R_{\mathrm{TF}}$ and $\tilde{f}\propto fR^3_{\mathrm{TF}}$
depends on $n(x)$ and $T$ (since $R_{\mathrm{TF}}\propto [\mu_0(T)]^{\frac{1}{p}}$).  Also, we have absorbed the contributions due to $g$ as well as the background contributions into the terms denoted by $H^{\pm}(T)$.  These terms also include the noncritical background contributions which are not described by (\ref{cT}) and come from the spatial region $R_{\mathrm{c}}<r<R_{\mathrm{TF}}$.  

Equations analogous to (\ref{Capprox}) and (\ref{Capprox2}) have been used to determine the exponent $\alpha$ in high-precision measurements done on large sample sizes of superfluid $^4$He where gravity effects give rise to a position-dependent transition temperature $T_{\mathrm{c}}(r)$~\cite{Heexpt2,Ahlers1,Ahlers2}.  In these studies, the reduced temperature $t(r)$ varied by only a small amount over the whole sample. This meant that one could be in a range of temperatures where the entire system was in the critical region and furthermore, in a single phase (i.e., superfluid \textit{or} normal).  Consequently, experiments could also extract information about the scaling amplitudes $f$ and $g$ from measurements of the bulk specific heat. In contrast, in trapped gases, $t(r)$ varies significantly over the gas since the density decreases rapidly away from the center.  Thus, the bulk specific heat of a trapped gas will always contain a contribution from the noncritical region at the edge of the gas.  Furthermore, for $t(0)<0$, one can never realize a region at finite temperatures where the entire gas is in the superfluid phase.    For these reasons, it is not possible to obtain detailed information about the scaling amplitudes $f$ and $g$ in trapped gases.

Nonetheless, one can still use (\ref{Capprox}) and (\ref{Capprox2}) to deduce the scaling behavior $C \propto |t|^{-\alpha'}$ (where $\alpha'$ is the exponent for the trapped gas) of the specific heat in the critical region as long as the changes in $H^{\pm}(T)$ and $\mu_0(T)$ are small over the range of absolute temperatures $T$ probed.  As with $^4$He, this requirement is easily realized since noncritical background contributions depend analytically on temperature. (The chemical potential of a unitary Fermi gas is also a smooth, slowly varying function of temperature~\cite{Bulgac07}).  What is different in inhomogenenous trapped gases is that the noncritical background terms $H^{\pm}(T)$ will almost certainly dominate over the critical contributions to the bulk specific heat if the volume $\propto R^3_{\mathrm{c}}$ of the critical region is much smaller than the volume $\propto R^3_{\mathrm{TF}}$ of the gas.  In this case, it would be difficult to experimentally resolve scaling behavior.  A related issue arises in weakly interacting trapped Bose gases, where the contribution to the shift in $T_{\mathrm{c}}$ (from its ideal gas value) due to noncritical background terms dominates over critical contributions as a result of the smallness of ($R_{\mathrm{c}}/R_{\mathrm{TF}})^3$~\cite{Tomasik}.  

At $T=T_{\mathrm{c}}(0)$, the Ginzburg critical radius becomes $R_{\mathrm{c}} = R_{\mathrm{TF}}[Gi/(Gi+1)]^{\frac{1}{p}}$ and the fraction of atoms in the critical region is thus $\sim[Gi/(Gi+1)]^{\frac{3}{p}}$.  For a Bose gas with scattering length $a$ and thermal de Broglie wavelength $\lambda_T$, above $T_{\mathrm{c}}$, $Gi\sim (a/\lambda_T)^2\ll 1$~\cite{Damle96}.  This means that, for atoms confined in a harmonic trap ($p=2$), the fraction of Bose atoms that are in the critical region is of the order~\cite{Tomasik} $(a/\lambda_T)^3\ll 1$.  In contrast, for a Fermi gas at unitarity, the fact that $Gi\sim {\cal{O}}[1]$ means that $R_{\mathrm{c}}\sim {\cal{O}}[R_{\mathrm{TF}}]$ and a significant fraction of the gas is in the critical region when $t(0)\sim 0$.  This suggests that scaling behavior of the bulk specific heat might be observable for strongly interacting Fermi gases.    

We start by considering the scaling behavior in the normal phase, $T>T_{\mathrm{c}}(0)$.  Evaluating the integral in (\ref{Capprox2}) and expanding the result in powers of $t(0)$, the leading terms are~\cite{intnote}
\bea C \sim c_0+ c_1t(0) + c_2[t(0)]^{\frac{3}{p}-\alpha}\label{Ct},\eea
where $c_0$, $c_1$, and $c_2$ are constants of the same order of magnitude.   For the case of a harmonic trapping potential ($p=2$), the contribution to the specific heat coming from the third term in the above expression has also been obtained in Ref.~\cite{Damle96}  (as one can see by applying the Josephson scaling relation~\cite{Josephson} $\alpha = 2-3\nu$ to the results of Ref.~\cite{Damle96}, where $\nu$ is the critical exponent for the coherence length).  In addition, however, we find a term that depends linearly on the reduced temperature $t(0)$.  For harmonic traps, this term will dominate the specific heat of a trapped gas in the region about $T_{\mathrm{c}}(0)$.   

Below $T_{\mathrm{c}}(0)$, the situation is more complicated due to the inhomogeneity of the gas.  In particular, note that the rate of change with respect to $|t(0)|$ of the integrals in the first and second lines of (\ref{Capprox}) have opposite signs.  Assuming that $f_s$ and $f_n$ are both negative numbers (as is the case with $^4$He and superconductors), this means that while the slope of the specific heat still changes discontinuously at $T=T_{\mathrm{c}}(0)$, the peak in the specific heat may occur below this cusp.  Lowering the temperature below $T_{\mathrm{c}}(0)$, the specific heat may continue to increase until the first line in (\ref{Capprox}) dominates over the second.  The existence and size of the displacement of this peak from $T_{\mathrm{c}}(0)$ depends on the values of the amplitudes $f_{s},f_n, g_s$, and $g_n$.  A similar situation arises in studies of inhomogeneous $^4$He (see for instance, Fig.~2 in Ref.~\cite{Heexpt2}).  For these reasons, it is better to investigate the scaling behavior of the specific heat in the normal phase.

Equation~(\ref{Ct}) shows how the scaling behavior of the specific heat of a uniform superfluid in the critical region becomes strongly suppressed in a trapped gas due to the extra term $3/p$.  (For positive critical exponents, the larger the exponent, the weaker the cusp.) It also shows how critical behavior is more pronounced in a ``box-like" trapping potential: the higher the power of the potential, the smaller the critical exponent and the more the trapped gas behaves like a uniform superfluid where $C\propto |t|^{-\alpha}$.   This suggests that one way to enhance critical effects is to trap the gas in as box-like a potential as possible. To this end, we also note that $R'_{\mathrm{c}}$ tends to unity as $p \to \infty$, meaning that the fraction of atoms in the critical region is also larger in higher-order potentials.

\subsection{Compressibility in a trapped gas}

We now consider the behavior of the local compressibility $\kappa_T(r)$ close to the critical point.  The advantage of dealing with this quantity is that one can probe the \textit{local} compressibility directly by looking at the density profile $n(r)$ of the gas, instead of dealing with a trap averaged quantity such as the bulk specific heat.  In principle, this allows for more direct access to the critical region.  

Using (\ref{LDA}), one finds that the local compressibility $\kappa_T(r)\equiv n^{-2}(r)(\partial \mu[n(r)]/\partial n)^{-1}_T$ of a harmonically trapped ($V_{p=2}=m\omega^2/2$) gas is related to the slope of the density profile by~\cite{Ho09,Zhou09} $\kappa_T(r) = -[mn^2(r)\omega^2r]^{-1}\partial n(r)/\partial r$.  (The constancy of temperature is enforced by the fact that it is independent of position.) Combining this with (\ref{kU}) and (\ref{Tcr}), in the superfluid phase we find (restoring $\hbar$)
\bea \frac{a^2_{ho}}{xn^{\frac{1}{3}}(x)}\frac{\partial n(x)}{\partial x}=\tilde{k}\left[\frac{x^2-|t(0)|}{1-x^2}\right]^{-\gamma} + \tilde{h}_{\kappa},
\label{Ktrap}\eea
for the normal region $r>R_s$ ($x>\sqrt{|t(0)|}$) of the gas.  Here, $a_{ho}\equiv \sqrt{\hbar/m\omega}$ is the harmonic oscillator length.  A similar equation applies to the superfluid region $r< R_s$ ($x<\sqrt{|t(0)|}$).  In (\ref{Ktrap}), $\tilde{k}\propto (\mu_0/\hbar\omega)k$ and $\tilde{h}_{\kappa}\propto (\mu_0/\hbar\omega)h_{\kappa}$ are dimensionless functions of $n$ and $T$. 

Mean-field theories of trapped Fermi superfluids (e.g., Ginzburg--Landau~\cite{Ho04} and BCS~\cite{Holland}) predict a discontinuity in the compressibility and hence, a ``kink" in the density profile (where $\partial n(x)/\partial x$ is discontinuous) at the local critical point $x=\sqrt{t(0)}$ using LDA~\cite{Holland,Ho04}.  In contrast, the LDA scaling relation (\ref{Ktrap}) predicts that the slope of the density profile diverges at $x=\sqrt{t(0)}$.  Neither of these features should be taken too seriously since LDA breaks down in the immediate vicinity of the local superfluid transition.   By looking at the region slightly away from the local critical point, however, detailed measurements of the density profile could in principle be used to determine the critical exponent $\gamma$.    We recall that the fact that the Ginzburg critical region is large means there is a large spatial region where (\ref{Ktrap}) is expected to be valid.  

Without detailed knowledge of the background function $\tilde{h}_{\kappa}(n,T)$ or the temperature dependence of the chemical potential $\mu_0(T)$, it is impossible to extract the critical exponent $\gamma$ by measuring the density profile $n(r)$ at a fixed temperature since $T_{\mathrm{F}} = T_{\mathrm{F}}(n)$ is position-dependent and the value of the density $n(r)$ can change rapidly over a small spatial region.  However, by measuring the left-hand side of (\ref{Ktrap}) for a fixed value of $x$ close to the local critical point and varying $t(0)$, for small changes in the absolute temperature $T$, it may be possible for experiments to resolve this critical exponent.  (The weak temperature dependence of the chemical potential $\mu_0(T)$ close to $T_{\mathrm{c}}$ in the unitary region means that $\tilde{k}$ is also weakly temperature dependent.)

\section{Implications for Experiments}
\label{Implicationssec}

We turn now to the question posed in the Introduction and try and understand whether any signatures of critical behavior can be seen in the measured specific heat~\cite{Kinast05,Luo08} and density profiles~\cite{Ketterle08} of a unitary Fermi gas.  

The experiments reported in Refs.~\cite{Kinast05,Luo08} observe an anomaly in the specific heat  at the putative transition temperature, but lack sufficient resolution for any strong conclusions to be drawn about the scaling behavior above $T_{\mathrm{c}}$.  It is thus important to estimate the range of temperatures above $T_{\mathrm{c}}$ where we expect the scaling law in (\ref{Ct}) to be valid and also the strength of the nonanalytic critical behavior (i.e., the size of $c_1$ and $c_2$ in (\ref{Ct})).  

The Ginzburg--Levanyuk number gives an estimate of the temperature range over which critical behavior manifests itself and we expect scaling behavior of the form given by (\ref{cU}) and (\ref{kU}) to be valid over much of this range.   To check this point, we look to experimental results in other strongly interacting \xy superfluids, $^4$He~\cite{Heexpt3} and cuprate superconductors, notably YBaCuO~\cite{Overend94}.  Experiments on $^4$He have exquisite control over the temperature and are able to probe the critical region in a tiny region about $T_{\mathrm{c}}$.  The most extensive experiments have measured the specific heat over eight decades: $10^{-10}<|t|<10^{-2}$~\cite{Heexpt3}.  These experiments confirm that scaling behavior of the form given by (\ref{cU}) is valid for \textit{at least} $|t| < 10^{-2}$.  In measurements of the specific heat of the cuprate superconductor YBaCuO, the same \xy scaling behavior has been observed in the region $|t|<10^{-1}$, \textit{larger} than the Ginzburg--Levanyuk number $Gi$.  An important reason why the specific heat in these strongly interacting \xy systems clearly exhibit the scaling behavior (\ref{cU}) over the entire Ginzburg critical region is that the coefficient $f$ is large compared to $g$ and $h_c$, and critical contributions to the specific heat dominate over noncritical background contributions (see, for instance, Ref.~\cite{Ahlers2}).  

Extrapolating these results to a strongly interacting Fermi gas close to unitary (which has a comparable critical region), the scaling law in (\ref{Ct}) should be valid over a large temperature range, at least $10^{-1}T_{\mathrm{c}}\simeq 0.02T_{\mathrm{F}}$. This is close to the resolution in current experiments which, for entropy $S$, we estimate to be around $\Delta T \sim (k_{\mathrm{B}}\Delta E/SE_{\mathrm{F}})T_{\mathrm{F}} \sim 0.01T_{\mathrm{F}}$ from the data in Ref.~\cite{Luo08}.  This suggests that these experiments are on the cusp of being able to resolve the linear scaling behavior predicted in (\ref{Ct}) above $T_{\mathrm{c}}(0)$, assuming that background effects do not overwhelm contributions from critical behavior.

The observability of the scaling predicted by (\ref{Ct}) is, of course, dependent on the size of the coefficients $c_1$ and $c_2$.  If these are small relative to the background contributions, then critical scaling may not be seen in experiments.   However, as pointed out in Sec.~\ref{SHsec}, the large size of the Ginzburg critical region means that a significant fraction of the gas is in the critical region when $T\sim T_{\mathrm{c}}(0)$, in contrast to the situation in weakly interacting Bose gases.  This means that as long as the amplitude $f$ in the scaling law (\ref{cU}) for the \textit{homogeneous} gas is not too small compared to background terms, then the amplitudes $c_1$ and $c_2$ for the \textit{inhomogeneous} trapped gas will not be too small either.  We cannot say anything about the magnitude of $f$ compared to $h_c(n,T)$ since the ratio of these quantities is nonuniversal and will be particular to the unitary Fermi gas.  However, a small value of $f$ (such that critical behavior is washed out by background effects) would be in sharp contrast to other strongly interacting \xy systems, where specific heat scaling is clearly evident.  

Finally, we recall that the peak in the specific heat may be situated below the transition temperature $T_{\mathrm{c}}(0)$.  The scaling behavior predicted by (\ref{Ct}) thus cannot be extended right down to the specific heat peak.  Equation~(\ref{Ct}) is only valid for $T$ greater than the superfluid transition temperature $T_{\mathrm{c}}(0)$, where the specific heat exhibits a (likely) weak nonanalyticity.  However, since the superfluid region grows quite fast in a trapped gas ($R_s=R_{\mathrm{TF}}|t(0)|^{\frac{1}{p}}$) and $f_s\simeq f_n$~\cite{Guillou}, from (\ref{Capprox}), we expect that the peak in the specific heat will be very close to $T=T_{\mathrm{c}}(0)$.  Asymptotically, as $p\rightarrow \infty$, the peak moves closer to $T=T_{\mathrm{c}}(0)$, giving rise to the expected nonanalytic peak in the specific heat (lambda curve) characteristic of a uniform superfluid.  

Turning now to our prediction (\ref{Ktrap}) for the density profile in the vicinity of the critical point, we note that until very recently, no experiment on a trapped Fermi gas close to unitary has observed any feature in the density profile indicating the presence of a superfluid component.  This suggests that the value of $\tilde{k}$ is very small and the divergence in the compressibility is very weak.  Note that this would be consistent with the situation in $^4$He, where the divergence of the isothermal compressibility is also weak~\cite{Vinen}.  

Recent detailed studies of the density profile of a unitary Fermi gas carried out at MIT do reveal a feature in the density profile (see Fig.~48 in Ref.~\cite{Ketterle08}).  However, the resolution in these experiments is not yet good enough--and the effect is too small--to be able to measure the critical exponent $\gamma$.

\section{Conclusions}

In summary, we have analysed the width of the critical region in a Fermi gas through the BCS-BEC crossover and discussed signatures of critical behavior in the specific heat.  We find that close to unitarity, the critical region is \textit{very} wide--wider than clean cuprate superconductors and comparable to superfluid $^4$He.  Inside the Ginzburg critical region, any method for calculating thermodynamic quantities that relies on an expansion in powers of fluctuations of the order parameter is intrinsically unreliable and cannot be expected to produce quantitatively accurate results. 

This is an important point to emphasize in light of recent comments~\cite{He07} in the literature concerning unphysical features predicted by various Gaussian fluctuation theories, notably a double-valuedness of the order parameter close to $T_{\mathrm{c}}$~\cite{Haussmann07,Fukushima07,note}.   This pathology is well-known in the Bose gas literature (see, for instance, Refs.~\cite{ShiGriffin,Holzmann03}) and directly reflects the break down of perturbation theory in the critical region.  Indeed, this double-valuedness occurs through the BCS-BEC crossover over a temperature range of the order $Gi$ about $T_{\mathrm{c}}$ (see Fig.~1(a) in Ref.~\cite{Fukushima07}).  One can thus interpret the numerical results reported in Refs.~\cite{Haussmann07,Fukushima07} as ``empirical" evidence of a large critical region close to unitarity.  Further evidence of a large critical region close to unitarity was found in Ref.~\cite{Strinati02}, where a diagrammatic analysis showed that diagrams that are subleading for $T\gg T_{\mathrm{c}}$ exceed leading ones in a wide region ${\cal{O}}[T_{\mathrm{c}}]$ above $T_{\mathrm{c}}$.  These results emphasize that the \textit{only} quantitatively reliable analytic method in the critical region is the renormalization group~\cite{Bijlsma96,Ohashi05,Diehl07,Gubbels08}.   

Despite the fact that the width of the critical region of a unitary Fermi gas is of the order unity, signatures of critical behavior in trapped gases are suppressed by their inhomogeneity.  For harmonic traps, we find that the bulk specific heat is \textit{linear} in the reduced temperature close to $T_{\mathrm{c}}$.  Furthermore, the strength of the critical behavior is diminished in traps where the bulk specific heat includes contributions from the noncritical region at the edge of the gas.  These factors likely explain why experiments have failed to observe a clear lambda-like curve in the specific heat, although a distinct feature is observed close to $T_{\mathrm{c}}$~\cite{Kinast05,Luo08}.  

We have also considered a scheme to measure critical behavior in the compressibility by looking at the density profile.  This scheme has the advantage that one can restrict the measurement to the critical region since it involves a local rather than bulk thermodynamic quantity.   However, the nonanalytic contributions to the compressibility seem to be small in comparison with noncritical background contributions, meaning that experiments will need good resolution to be able to measure the critical exponent for the compressibility.  Nonetheless, a feature has already been observed in the density profile of a trapped unitary Fermi gas below the superfluid transition~\cite{Ketterle08} and future experiments may be able to explore this in greater detail.

\section{Acknowledgements}  

I am grateful to the organizers for their support during the workshop ``Ultracold Atoms and Quark-Gluon Plasmas" at 
the Niels Bohr International Academy and NORDITA, 2008, where this work was initiated.  
I also want to thank Lev~Pitaevskii for helpful discussions.  

\appendix
\section{Validity of local density approximation}
\label{LDAsec}

Close to the superfluid transition, the local density approximation (LDA) used to obtain the results in Sec.~\ref{SHsec} should be valid as long as the length scale $R_{\xi}$ over which the local (Ginzburg--Landau) coherence length $\xi(r,T)$ varies significantly is much larger than the coherence length $\xi(r=0,T)$ at the trap center: $\xi(0,T)/R_{\xi}\ll 1$. This is analogous to the usual finite-size criterion for a uniform superfluid confined in a box with sides of length $L$ that requires $\xi/L\ll 1$ in order to be able to ignore finite-size effects~\cite{Gasparini08}.  For a gas confined in a harmonic potential $V = m\omega^2r^2/2$, the local coherence length is $\xi(r,T)\propto |T-T_{\mathrm{c}}(r)|^{-\nu}\propto (1-r^2/R^2_{\xi})^{-\nu}$, where $\nu\simeq 0.67$, and $R_{\xi} \equiv R_{\mathrm{TF}}\sqrt{|t(0)|}$.   The Thomas--Fermi radius is (apart from an interaction renormalization factor of the order unity and restoring $\hbar$)~\cite{Trentoreview} $R_{\mathrm{TF}} = \sqrt{\hbar/m\omega}(24 N)^{1/6}$, where $N$ is the atom number.  Making use of the fact that the $T=0$ coherence length in a homogeneous unitary Fermi gas is~\cite{Pistolesi94,Engelbrecht97} $\sim k^{-1}_{\mathrm{F}}$, we take the coherence length at the trap center to be $\xi(r=0,T) \sim k^{-1}_{\mathrm{F}}(r=0)|t(0)|^{-\nu}$, where $k_{\mathrm{F}}(r=0)  = \sqrt{2m E_{\mathrm{F}}/\hbar^2}$ and $E_{\mathrm{F}} = \hbar\omega(3N)^{1/3}$ is the bulk Fermi energy of a trapped gas.  Thus, the condition $\xi(0) \ll R_{\xi}$ is satisfied when $|t(0)|\ll(24N)^{-2/(3+6\nu)}$.  In typical experiments ($N\sim 10^5$), this means that LDA is valid except in a small region $|t(0)|<10^{-2}$ about $T_{\mathrm{c}}$.  Note that this is equal to the estimate given in Ref.~\cite{Damle96} for a weakly interacting dilute Bose gas.  For smaller values of $|t(0)|$, when $\xi(0,T)\sim R_{\xi}$, ``finite-size" effects will become important.


\begin{thebibliography}{99} 
\bibitem{Trentoreview} S.~Giorgini, L.~P.~Pitaevskii, and S.~Stringari, Rev. Mod. Phys. \textbf{80}, 1215 (2008).
\bibitem{Chen05} Q.~Chen, J.~Stajic, and K.~Levin, Phys. Rev. Lett. \textbf{95}, 260405 (2005).
\bibitem{Haussmann07} R. Haussmann, W. Rantner, S. Cerrito, and W. Zwerger, Phys. Rev. A \textbf{75}, 023610 (2007)
\bibitem{Hu07} H.~Hu, P.~D.~Drummond, and X.-J.~Liu, Nat. Phys. \textbf{3}, 469 (2007); H.~Hu, X.-J.~Liu, and P. D. Drummond, Phys. Rev. A \textbf{77}, 061605(R) (2008). 
\bibitem{Bulgac07} A.~Bulgac, J.~E.~Drut, and P.~Magierski, Phys. Rev. Lett. \textbf{99}, 120401 (2007).
\bibitem{Haussmann08} R.~Haussmann and W.~Zwerger, Phys. Rev. A \textbf{78}, 063602 (2008).
\bibitem{Kinast05} J.~Kinast, A.~Turlapov, J.~E.~Thomas, Q.~Chen, J.~Stajic, and K.~Levin, Science \textbf{307}, 1296 (2005).
\bibitem{Luo07} L.~Luo, B.~Clancy, J.~Joseph, J.~Kinast, and J.~E.~Thomas, Phys. Rev. Lett. \textbf{98}, 080402 (2007).
\bibitem{Luo08} L.~Luo and J.~E.~Thomas, J. Low Temp. Phys. \textbf{154}, 1 (2009).   
\bibitem{LPSM}  E.~M.~Lifshitz and L.~P.~Pitaevskii, \textit{Statistical Physics, Part 2} (Butterworth-Heinemann, Oxford, 2002).
\bibitem{Zinn} J.~Zinn-Justin, \textit{Quantum Field Theory and Critical Phenomena} (Clarendon Press, Oxford, 1996). 
\bibitem{Kadanoff}
L.~P.~Kadanoff, W.~G\"otze, D.~Hamblen, R.~Hecht, E.~A.~S.~Lewis, V.~V.~Palciauskas, M.~Rayl, and J.~Swift, Rev. Mod. Phys. \textbf{39}, 395 (1967).
\bibitem{Cobb} C.~J.~Lobb, Phys. Rev. B \textbf{36}, 3930 (1987).
\bibitem{Deutscher} G.~Deutscher, in \textit{Novel Superconductivity}, S.~A.~Wolf and V.~Z.~Kresin, eds. (Plenum, New York, 1987). 
\bibitem{Damle96} K.~Damle, T.~Senthil, S.~N.~Majumdar, and S.~Sachdev, Europhys. Lett. \textbf{36}, 7 (1996). 
\bibitem{Giorgini96} S.~Giorgini, L.~P.~Pitaevskii, and S.~Stringari,  Phys. Rev. A \textbf{54}, R4633 (1996).
\bibitem{Donner07} T.~Donner, S.~Ritter, T.~Bourdel, A.~\"Ottl, M.~K\"ohl, and T.~Esslinger, Science \textbf{315}, 1556 (2007). 
\bibitem{Heexpt} M.~J.~Buckingham and W.~M.~Fairbank, \textit{Progress in Low Temperature Physics}, edited by J.~C.~Gorter (North-Holland Publishing Company, Amserdam, 1961), Vol.~3.
\bibitem{SadeMelo93} C.~A.~R.~ S\'a de Melo, M.~Randeria, and J.~R.~Engelbrecht, Phys. Rev. Lett  \textbf{71}, 3202 (1993).
\bibitem{BCSBEC} A.~J.~Leggett, J. Phys. Colloques \textbf{41}, C7 (1980); D.~M.~Eagles, Phys. Rev. \textbf{186}, 456 (1969).  
\bibitem{NSR} P.~Nozi\`eres and S.~Schmitt-Rink, J. Low Temp. Phys. \textbf{59}, 195 (1985). 
\bibitem{RanderiaTalk} M.~Randeria, talk given at KITP Program: Strongly Correlated Phases in Condensed Matter and Degenerate Atomic Systems, Santa Barbara, (2007).  
\bibitem{Pistolesi94} F.~Pistolesi and G.~C.~Strinati, Phys. Rev. B \textbf{49}, 6356 (1994); Phys. Rev. B \textbf{53}, 15168 (1996).
\bibitem{Engelbrecht97} J.~R. Engelbrecht, M.~Randeria, and C.~A.~R. S\'a de Melo, Phys. Rev. B  \textbf{55}, 15153 (1997).
\bibitem{Heexpt2} J.~A.~Lipa and T.~C.~P.~Chui, Phys. Rev. Lett. \textbf{51}, 2291 (1983).
\bibitem{Ahlers1} G.~Ahlers, Phys. Rev. \textbf{171}, 275 (1968).
\bibitem{Ahlers2} G.~Ahlers, Phys. Rev. A \textbf{3}, 696 (1971). 
\bibitem{LLSM} L.~D.~Landau and E.~M.~Lifshitz, \textit{Statistical Physics} (Elsevier, Oxford, 2003), Sec.~146. 
\bibitem{noteGL} The constant $1/32\pi^2$ in (\ref{Gi}) is somewhat arbitrary and there are a number of definitions of $Gi$ in the literature, all with the same dependence on the Ginzburg--Landau coefficients but with different values of this prefactor (also depending on weather one is working in the normal or superfluid phase~\cite{LLSM}).  Our definition in (\ref{Gi}) is the same as that used in Ref.~\cite{Cobb} and we use this value in both the superfluid and normal phases for simplicity.                 
\bibitem{Gorkov} L.~P.~Gork'ov,  Sov. Phys. JETP \textbf{9}, 1364 (1959).  
\bibitem{ShiGriffin} H.~Shi and A.~Griffin, Phys. Rep. \textbf{304}, 1 (1998).  
\bibitem{Andrenacci03} N. Andrenacci, P. Pieri, and G. C. Strinati, Phys. Rev. B \textbf{68}, 144507 (2003).  
\bibitem{Pieri05} P.~Pieri and G.~C.~Strinati, Phys. Rev. B \textbf{71}, 094520 (2005).
\bibitem{Diener08} R.~B.~Diener, R.~Sensarma, and M.~Randeria, Phys. Rev. A \textbf{77}, 023626 (2008). 
\bibitem{Fukushima07} N.~Fukushima, Y.~Ohashi, E.~Taylor, and A.~Griffin, Phys. Rev. A \textbf{75}, 033609 (2007). 
\bibitem{Guillou} J.~C.~Le~Guillou and J.~Zinn-Justin, Phys. Rev. B \textbf{21}, 3976 (1980).
\bibitem{Guillou2} J.~C.~Le~Guillou and J.~Zinn-Justin, J.~Phys. Lett. \textbf{46}, L135 (1985).   
\bibitem{Ho09} T.-L.~Ho and Q.~Zhou, arXiv:0901.0018.  
\bibitem{Zhou09} Q.~Zhou, Y.~Kato, N.~Kawashima, and N.~Trivedi, arXiv:0901.0606.
\bibitem{Tomasik} P.~Arnold and B.~Tom\'{a}\v{s}ik, Phys. Rev. A \textbf{64}, 053609 (2001). 
\bibitem{intnote} Equation~(\ref{Ct}) has been derived by taking $\tilde{f}_n$ to be a constant.  One can readily confirm that any smooth function of position and temperature will not change this result, however.   We also note that the scaling form of Eq.~(\ref{Ct}) is independent of the $t(0)$-dependence of the upper limit of integration, $R'_c$.  
\bibitem{Josephson} B.~D.~Josephson, Phys. Lett. \textbf{21}, 608 (1966).
\bibitem{Ho04} T.-L.~Ho, Phys. Rev. Lett. \textbf{92}, 090402 (2004).
\bibitem{Holland} M.~Holland, S.~J.~J.~M.~F.~Kokkelmans, M.~L.~Chiofalo, and R.~Walser, Phys. Rev. Lett. \textbf{87}, 120406 (2001).
\bibitem{Ketterle08} W.~Ketterle and M.~W.~Zwierlein, in \textit{Proceedings of the International School of Physics ``Enrico Fermi" - Course CLXIV ``Ultra-Cold Fermi Gases", Varenna, June 2006}, edited by M.~Inguscio, W.~Ketterle, and C.~Salomon (IOS Press, Amsterdam, 2008). 
\bibitem{Heexpt3} J.~A.~Lipa, J.~A.~Nissen, D.~A.~Stricker, D.~R.~Swanson, and T.~C.~P.~Chui, Phys. Rev. B \textbf{68}, 174518 (2003). 
\bibitem{Overend94} N.~Overend, M.~A.~Howson, and I.~D.~Lawrie, Phys. Rev. Lett. \textbf{72}, 3238 (1994).
\bibitem{Vinen} W.~F.~Vinen and J.~M.~Vaughan, J. Phys. Colloques \textbf{31}, C3-29 (1970).
\bibitem{He07} Y.~He, C.-C.~Chien, Q.~Chen, and K.~Levin, Phys. Rev. B \textbf{76}, 224516 (2007).
\bibitem{note} Ref.~\cite{He07} predicts a second-order phase transition.  The omission of long-wavelength phase fluctuations (phonons) avoids the problems of being in the critical region.  
\bibitem{Holzmann03}  M.~Holzmann and G.~Baym, Phys. Rev. Lett. \textbf{90}, 040402 (2003). 
\bibitem{Strinati02} G.~C.~Strinati, P.~Pieri, and C.~Lucheroni, Eur. Phys. J. B \textbf{30}, 161 (2002).    
\bibitem{Bijlsma96}  M.~Bijlsma and H.~T.~C.~Stoof, Phys. Rev. A \textbf{54}, 5085 (1996).
\bibitem{Ohashi05} Y.~Ohashi, J. Phys. Soc. Japan \textbf{74}, 2659 (2005). 
\bibitem{Diehl07} S.~Diehl, H.~Gies, J.~M.~Pawlowski, and C.~Wetterich, Phys. Rev. A \textbf{76}, 021602(R) (2007). 
\bibitem{Gubbels08}  K.~B.~Gubbels and H.~T.~C.~Stoof, Phys. Rev. Lett. \textbf{100}, 140407 (2008).
\bibitem{Gasparini08} F.~M.~Gasparini, M.~O.~Kimball, K.~P.~Mooney, and M.~Diaz-Avila, Rev. Mod. Phys. \textbf{80}, 1009 (2008).
\end{thebibliography}
\end{document}